\documentclass{iau}
\usepackage{graphicx}
\usepackage{natbib}
\usepackage{hyperref}
\hypersetup{
    final=true,
    pageanchor=true,
    colorlinks=true,
    breaklinks=true,
    linkcolor=blue,
    citecolor=blue,
    urlcolor=blue,
    pdfpagemode=UseNone,
    pdftitle={Statistical characteristics of horizontal proper motions in the
        vicinity of pores},
    pdfauthor={M. Verma and C. Denker},
    pdfsubject={Solar Physics}}
\usepackage{nicefrac}
\usepackage{amssymb}
\usepackage[usenames, dvipsnames]{color}
\usepackage[modulo,switch]{lineno}
\newcommand{\arcdeg}{\mbox{$^{\circ}$}}

\title[] 
{Statistical characteristics of horizontal proper motions in the vicinity of
pores}

\author[Meetu Verma \& Carsten Denker]   
{Meetu Verma \and Carsten Denker}

\affiliation{Leibniz-Institut f\"ur Astrophysik Potsdam, An der Sternwarte 16,
14482 Potsdam, Germany\\ 
email: {\tt mverma@aip.de and cdenker@aip.de}}

\pubyear{2012}
\volume{294}  
\pagerange{119--126}
\jname{Solar and Astrophysical Dynamos and Magnetic Activity}
\editors{A.G. Kosovichev, E.M. de Gouveia Dal Pino, \& Y.Yan, eds.}
\begin{document}

\maketitle

\begin{abstract} Movement and coalescence of magnetic elements could explain the
evolution and growth of pores. There have been numerous studies focusing on flow
fields in and around individual pores. We have undertaken a systematic study of
the statistical properties of such flows. Data of the \textit{Hinode} Solar
Optical Telescope offer an opportunity for this type of research, because of the
uniform data quality and absence of seeing so that pores can directly be
compared in different environments and at various stages of their evolution. We
analyzed about 220 time-series of G-band images using local correlation
tracking. The thus computed flow maps make up a database, which covers various
scenes on the solar surface. We use an isolated pore to illustrate the
statistical parameters collected for further statistical analyses, which
include information about morphology, horizontal flows, evolutionary stage
(young, mature, or decaying), complexity of the surrounding magnetic field, and
proximity to sunspots or cluster of G-band bright points.

\end{abstract}

\firstsection 

\section{Introduction}
Photospheric flow fields play an important role in the formation,
evolution, and decay of solar magnetic features. Our Local Correlation Tracking
(LCT) algorithm \citep{Verma2011} is based on ideas put forward by
\citet{November1988}. The algorithm was adapted to subimages with sizes of 32
$\times$ 32 pixels (2560~km $\times$ 2560~km), so that structures with
dimensions smaller than a granule contribute to the correlation signal. Because
cross-correlation techniques are sensitive to strong intensity gradients, a
high-pass filter was applied to the entire image, suppressing gradients related
to structures larger than granules. The high-pass filter is implemented as a
Gaussian with a FWHM of 15 pixels (1200 km).

\section{Statistical properties of photospheric flow fields around pores}
Starting point for identifying pores is a time averaged (one-hour) G-band image,
which has been corrected for limb darkening. Some additional smoothing and
subsequent thresholding at 90\% of the quiet-Sun intensity $I_0$ yields a binary
mask marking all dark features in the image. The intensity threshold roughly
corresponds to the transition from penumbra to umbra in a sunspot. Morphological
closing smoothes boundaries and eliminates isolated dark pixels. Only features
with a size between 1~Mm$^2$ and 225~Mm$^2$ are kept. These are conservative
limits for magnetic knots and sunspots, respectively. The next criterion is that
a pore must have a dark core, i.e., pixels with an intensity lower than
0.7$I_0$ should cover at least 0.5 Mm$^2$. Nevertheless, rudimentary or orphan
penumbrae are sometimes erroneously classified as pores. However, in almost all
cases, these features are larger than 20~Mm$^2$, have an intensity contrast
lower than 20\%, and possess a high positive kurtosis. Thus, these feature can
also be eliminated from the binary mask. 

\begin{figure}
\begin{center}
 \includegraphics[width=2.4in]{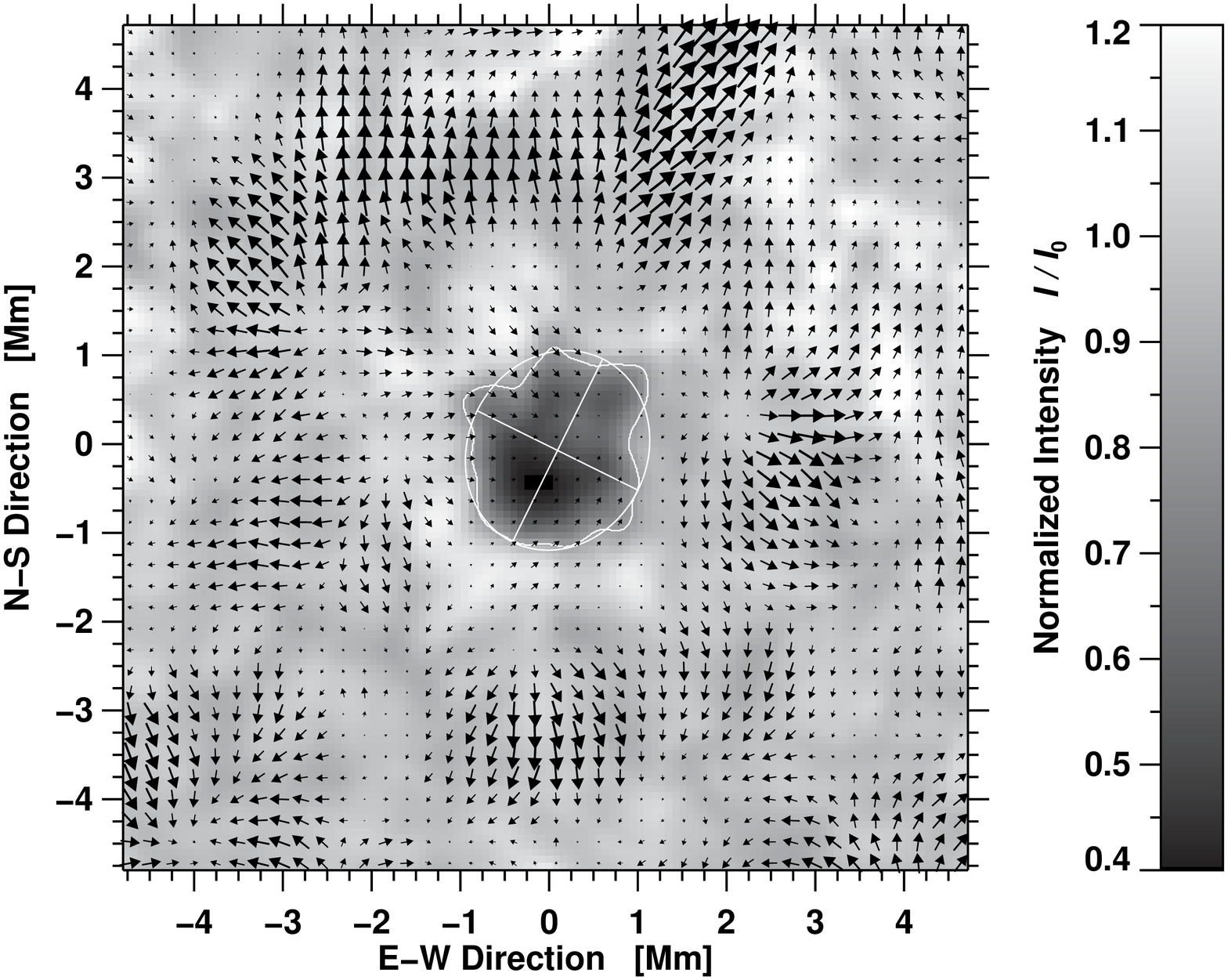} 
 \includegraphics[width=2.84in]{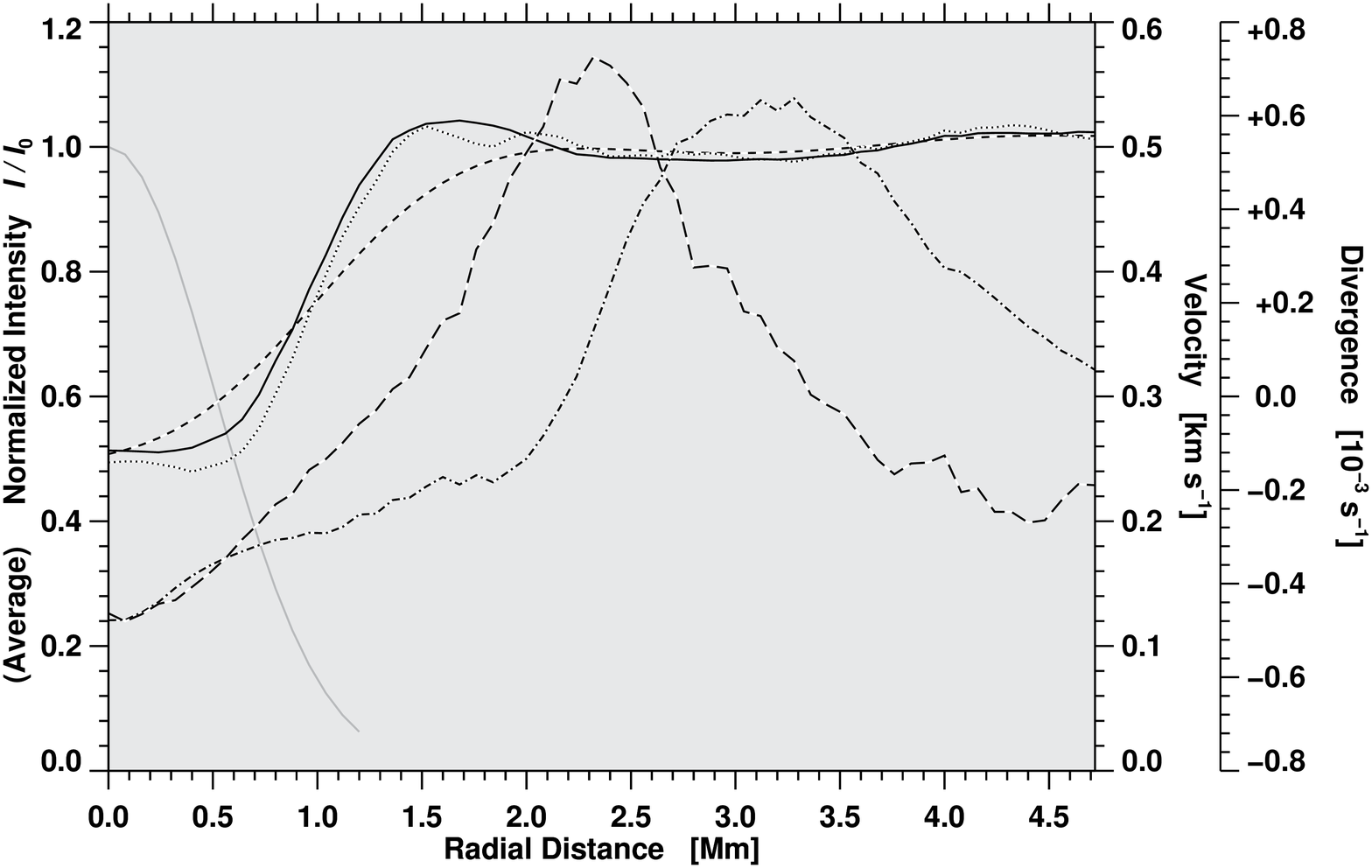} 
 \caption{Average (60-minute) G-band image (\textit{left}) of an isolated pore
observed on 2007 January~17. The arrows indicate magnitude and direction of the
horizontal proper motions. An arrow with the length of the grid spacing
indicates a speed of 0.5~km~s$^{-1}$. Azimuthally averaged profiles
(\textit{right}) of the G-band intensity (\textit{dotted}), the time-averaged
G-band intensity (\textit{solid}), the horizontal flow velocities
(\textit{dash-dotted}), and the divergence (\textit{long-dashed}). The
\textit{dashed} curve is the G-band intensity profile after convolving with a
Gaussian kernel (\textit{gray}, FWHM $=1.2$~Mm).}
\label{fig1}
\end{center}
\end{figure}

Statistical properties of pores and flows can now be computed for each
contiguous region in the binary mask. Deriving average values for the flows is
very easy because the time-averaged G-band image exactly matches the LCT flow
maps. Standard tools for blob analysis \citep{Fanning2003} are used to derive
parameters describing the morphology of pores and the associated flow field. We
measured the area $A=3.0$~Mm$^2$ covered by the pore, the length $L=7.2$~Mm of
the circumference, the ratio $A / L = 0.41$~Mm, the minimum
$I_{\mathrm{min}}=0.39$ and the mean $I_{\mathrm{mean}}=0.61$ intensities, the
average flow speed $v_{\mathrm{mean}}=0.16$~km~s$^{-1}$, the divergence $\nabla
v = -1.8 \times 10^{-4}$~s$^{-1}$, and the vorticity $\nabla \times v =
-1.1\times 10^{-6}$~s$^{-1}$. Fitting an ellipse to each pore yields among other
parameters the eccentricity $e = 0.47$ and the angle $\alpha = 62.1$\arcdeg\ of
the major axis with the horizontal (all values refer to the pore shown in
Fig.~\ref{fig1}). Radial averages beyond the pore's boundary are only computed,
if there are segments covering at least 180\arcdeg\ in azimuth which are free of
any other dark features. An example of blob analysis and azimuthally averaged
profiles is presented in Fig.~\ref{fig1}. A striking feature in the intensity
map is a bright ring around the pore and an annular structure of positive
divergence, which envelops the bright intensity ring like an onion peel. The
respective divergence centers in the annular structure are related to exploding
granules. Inflows inside the pore and outflows in the exterior are commonly
found in pores, and the radial velocity profile has a slow gradient inside pore.
The detailed results of the statistical study are deferred to an upcoming
publication.

\begin{acknowledgements}
\noindent \textit{Hinode} is a Japanese mission developed and launched by
ISAS/JAXA, with NAOJ as domestic partner and NASA and STFC (UK) as international
partners. It is operated by these agencies in co-operation with ESA and NSC
(Norway). MV thanks the German Academic Exchange Service (DAAD) for its support
in the form of a PhD scholarship. CD was supported by grant DE 787/3-1 of the
German Science Foundation (DFG).
\end{acknowledgements}

\end{document}